\documentclass[3p,times,twocolumn]{elsarticle}

\usepackage{ecrc}

\volume{00}

\firstpage{1}

\journalname{Nuclear Physics B Proceedings Supplement}

\runauth{O. Benhar}

\jid{nuphbp}

\jnltitlelogo{Nuclear Physics B Proceedings Supplement}

\usepackage{amssymb}

\usepackage[figuresright]{rotating}

\def\lsim{\buildrel < \over {_{\sim}}}

\def\beq{\begin{equation}}
\def\eeq{\end{equation}}
\def\be{\begin{eqnarray}}
\def\ee{\end{eqnarray}}
\begin{document}

\begin{frontmatter}



\dochead{}

\title{Modeling neutrino-nucleus interactions. Do we need a new paradigm?}


\author{Omar Benhar}

\address{INFN, Sezione di Roma \\
Dipartimento di Fisica, ``Sapienza" Universit\`a di Roma \\
I-00185 Roma, Italy}

\begin{abstract}
The availability of the double-differential charged-current neutrino cross section, measured by the 
MiniBooNE collaboration using a carbon target, allows for a systematic
comparison of nuclear effects in quasi-elastic electron and neutrino scattering. The results 
of theoretical studies based on the impulse approximation scheme and state-of-the-art models of the nuclear spectral functions 
suggest that the electron cross section and the flux averaged neutrino cross sections corresponding to the
same target and seemingly comparable kinematical conditions can not be described within the same theoretical scheme using the value of 
the nucleon axial mass obtained from deuterium measurements. I analyze the
assumptions underlying the treatment of electron scattering data, and argue that the development of 
a new {\em paradigm}, suitable for application to processes in which the lepton kinematics is not 
fully determined, will be required.
\end{abstract}


\end{frontmatter}


\section{Introduction}
\label{intro}
The data set of Charged Current Quasi Elastic (CCQE) events recently released by the MiniBooNE
collaboration \cite{BooNECCQE}  provides an unprecedented opportunity to carry out a systematic
study of the double differential cross section of the process,
\beq
\nu_\mu + ^{12}\mkern -5mu C \rightarrow \mu^- + X \ ,
\eeq
averaged over the neutrino flux.  


The charged current elastic neutrino-nucleon process is described in terms of three
form factors. The proton $(p)$ and neutron $(n)$ vector form factors, $F_1^{p,n}(Q^2)$ and
$F_2^{p,n}(Q^2)$ ($Q^2= -q^2$, $q$ being the four-momentum
transfer), have been precisely measured
up to large values of $Q^2$ in electron-proton and electron-deuteron scattering experiments,
respectively (for a recent review, see, e.g., Ref.\cite{VFF}). The $Q^2$-dependence of
the nucleon axial form
factor $F_A(Q^2)$, whose value at $Q^2=0$ can be extracted from neutron $\beta$-decay
measurements, is  generally assumed to be of dipole form and parametrized in terms of  the so
called axial mass $M_A$:
\beq
F_A(Q^2) = g_A \ \left( 1 + Q^2/M_A^2 \right)^{-2} \ .
\eeq
The world average of the measured values of the axial mass, mainly
obtained from low statistics experiments carried out using deuterium targets, turns out
to be  $M_A~=~1.03~\pm~0.02$~GeV \cite{bernard,bodek,nomad},
while the analyses performed by the K2K \cite{K2K} and MiniBooNE \cite{BOONE}
collaborations using oxygen and carbon targets, respectively, yield
$M_A \sim 1.2 \div 1.35 \ {\rm GeV}$.

It would be tempting  to interpret the large value of $M_A$ reported by MiniBoonNE
and K2K as an {\em effective} axial mass, modified by nuclear effects not included in the oversimplified
Fermi gas model employed in data analysis. However, most existing models of nuclear
effects (for recent reviews see Ref.\cite{nuint09}) fail to support this explanation,
suggested  by the authors of Ref.\cite{BOONE}, a prominent exception being the model of Ref.\cite{martini}.

Obviously, a fully quantitative description of the electron-scattering cross section, driven by the known
vector form factors, is a prerequisite for the understanding of the axial vector contribution to the
CCQE neutrino-nucleus cross section.

Over the past two decades, the availability of a large body of experimental data has triggered
the development of advanced theoretical descriptions of the nuclear electromagnetic response.
The underlying scheme, based on nuclear many-body theory and realistic nuclear hamiltonians,
relies on the premises that i) the lepton kinematics is fully determined and ii) the elementary
interaction vertex can be extracted from measured proton and deuteron cross sections.

The above {\em paradigm} has been successfully applied to explain the electron-nucleus cross section
in a variety of kinematical regimes (for a recent review of the quasi-elastic sector see Ref.\cite{RMP}). 
However,  in view of the
uncertainties associated with the energy of the incoming beam, the identification of the reaction
mechanisms and the determination of the interaction vertex, its extension to the case of neutrino
scattering may not be straightforward.

The comparison between theoretical calculations and data, presented in Section \ref{enuA}, 
suggests that the measured electron- and neutrino-nucleus cross sections can not be explained within the 
same scheme. The difficulties associated with the description of the flux unfolded total cross section 
are discussed in Section \ref{paradigm}. Finally, in Section \ref{conclusions} I argue that the 
the {\em paradigm} succesfully employed to analyze electron-nucleus scattering is not suitable
for application to processes in which the lepton kinematics is not fully determined, and must 
be significantly modified. 


\section{Electron- and neutrino-nucleus interactions}
\label{enuA}

\begin{figure}[h!]
\includegraphics[width=0.95\columnwidth]{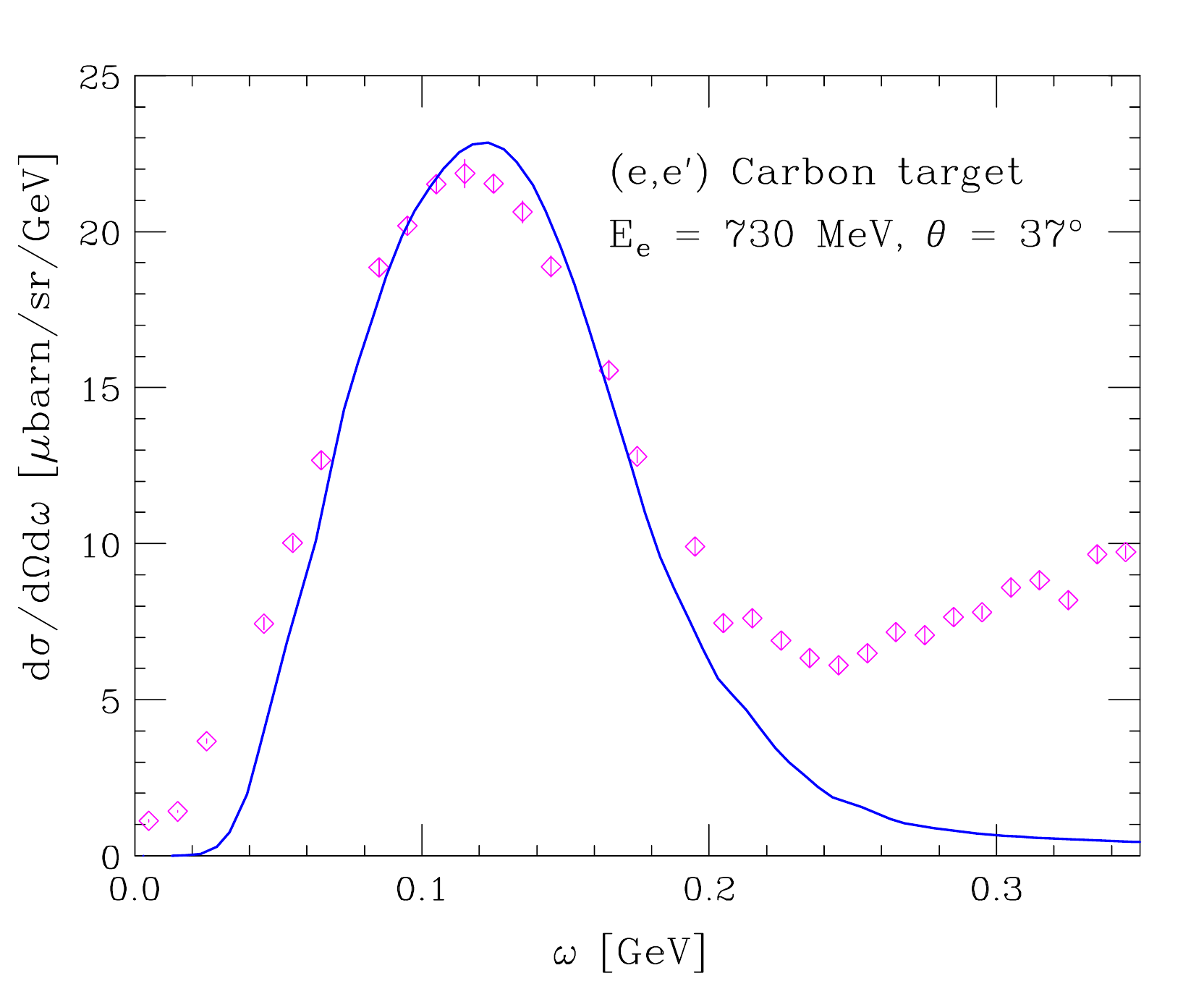}
\caption{Inclusive electron-carbon cross section at beam energy $E_e=$~730~MeV and electron scattering
angle $\theta_e=37^\circ$, plotted as a function of the energy loss $\omega$ \cite{PRL}. The data points are taken from
Ref.\cite{12C_ee}.}
\label{xsec_ee}
\end{figure}

Electron-nucleus scattering cross sections are usually analyzed at fixed beam energy, $E_e$, and
electron scattering angle, $\theta_e$, as a function of the electron energy loss $\omega$.
As an example, Fig. \ref{xsec_ee} shows the double differential cross section of the process
\beq
e + ^{12}\mkern -5mu C \rightarrow e^\prime + X \ ,
\eeq
at $E_e = 730$ MeV and $\theta_e = 37^\circ$, measured at MIT-Bates~\cite{12C_ee}.
The peak corresponding to quasi-elastic (QE) scattering, the bump at larger $\omega$, associated
with excitation of the $\Delta$-resonance, and the region in-between, where the cross section is mainly arising
from processes involving meson exchange currents (MEC), are clearly recognizable.
The three-momentum transfer $|{\bf q}|$ turns out to be nearly constant, its variation over the range
shown in the figure being $\lsim$ 5\%. As a consequence, the cross section of Fig.\ref{xsec_ee} can be
readily related to the linear  response of the target nucleus to a probe delivering momentum ${\bf q}$ and
energy $\omega$, defined as
\beq
\label{Sqw}
 S({\bf q},\omega) = \sum_n | \langle n | \sum_{{\bf k}} a^\dagger_{{\bf k}+{\bf q}} a_{{\bf k}} | 0 \rangle |^2
 \delta(\omega  + E_0 - E_n)\ .
 \eeq
In the above equation, $| 0 \rangle$ and $| n \rangle$ represent the target initial and final states, with
energies $E_0$ and $E_n$, respectively, while  $a^\dagger_{{\bf k} + {\bf q}}$ and $a_{{\bf k}}$ are the nucleon
creation and annihilation operators.

The magnitude of the momentum transfer,  $|{\bf q}|~\sim$~450~MeV,
is large enough to make the impulse approximation (IA) scheme, in which the nuclear response 
of Eq.(\ref{Sqw}) reduces to \cite{NUINT07}
\beq
S_{IA}({\bf q},\omega) = \int d^3k \ dE P_h({\bf k},E) P_p({\bf k}+{\bf q},\omega-E) \ ,
\label{S:IA}\eeq
safely applicable \cite{shape}. In  Eq.(\ref{S:IA}), the hole and particle spectral functions,  
$P_h(~{\bf k}~,~E~)$ and $P_p({\bf k}~+~{\bf q}~,~\omega~-~E)$, 
describe the energy and momentum distributions of the struck nucleon in the initial (hole) and
final (particle) states, respectively.

The solid line of Fig. \ref{xsec_ee} represents the results of a theoretical calculation of the QE contribution \cite{PRL}, carried 
out within the approach described in Refs.\cite{PRD,NPA} using the hole spectral function of Ref.\cite{PkE} and the recent 
parametrization of the vector form factors of Ref.\cite{bodek}. 
Final state interactions (FSI) between the struck nucleon and the recoiling  spectator system \cite{PRD}, 
whose main effect is a~$\sim 10$ MeV shift of the QE peak, have been also taken into account.

It is apparent that height, position and width of the QE peak, mostly driven by the energy and momentum 
dependence of the hole spectral function, are well reproduced.

\begin{figure}[ht]
\includegraphics[width=0.95\columnwidth]{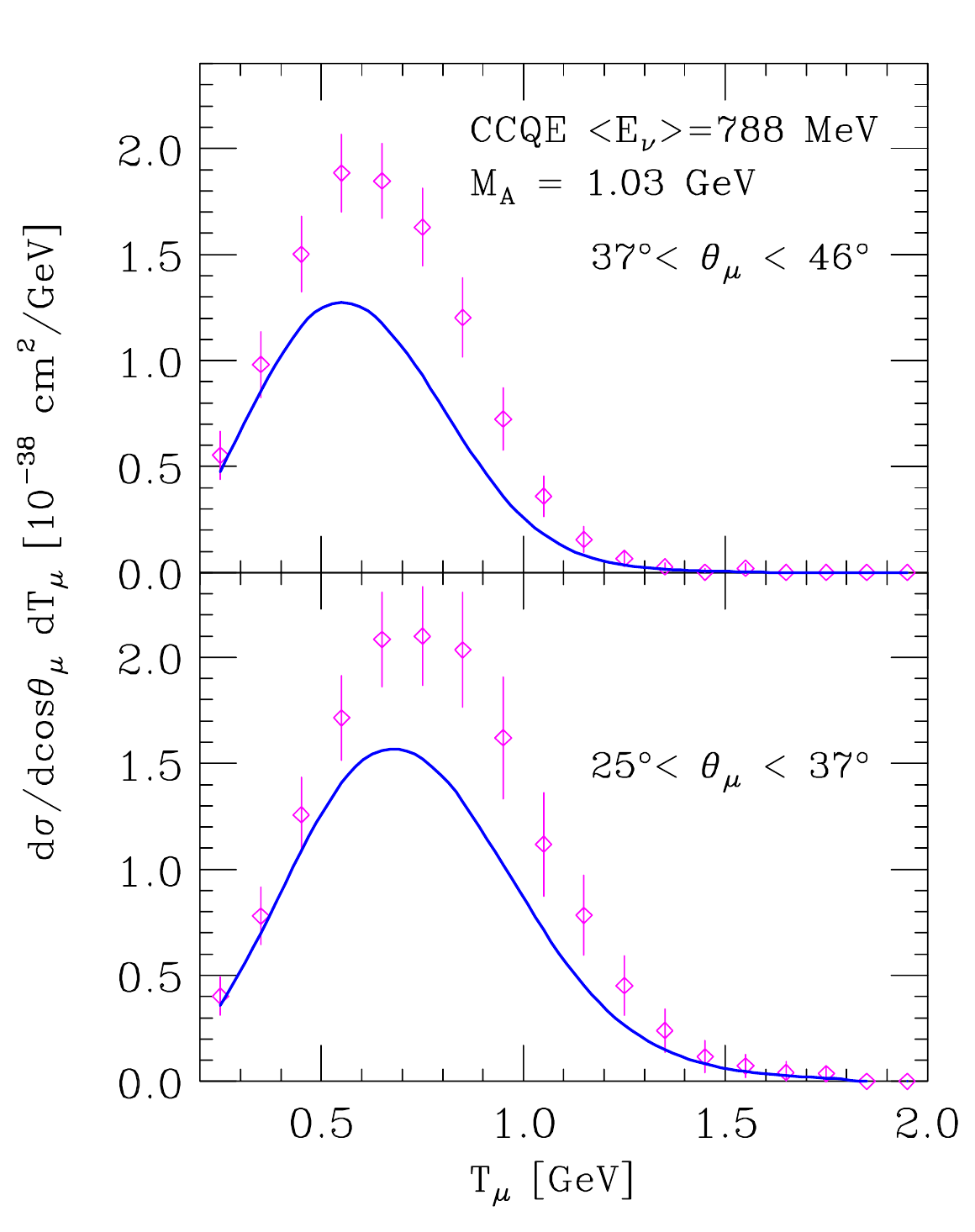}
\caption{Flux averaged double differential CCQE cross section measured by the MiniBooNE collaboration
\cite{BooNECCQE},  shown as a function of kinetic energy of the outgoing muon. The upper and lower panels correspond to
to different values of the muon scattering angle. Theoretical results have been obtained using the same spectral
functions and  vector form factors employed in the calculation of the electron scattering cross section of Fig. \ref{xsec_ee},
and a dipole parametrizaition of the axial form factor with $M_A=1.03$ MeV \cite{PRL}.}
\label{dsigma}
\end{figure}

Applying the same scheme employed to obtain the solid line of Fig. \ref{xsec_ee} to neutrino scattering
leads to the results shown in Fig. \ref{dsigma} \cite{PRL}. The data points represent the double differential CCQE cross
section averaged over the MiniBooNE neutrino flux, whose mean energy is $\langle~E_\nu~\rangle~=~788$~MeV,
plotted as a function of the kinetic energy of the outgoing muon, $T_\mu$, at different
values of the muon scattering angle $\theta_\mu$.
The solid lines show the results (integrated over the $\cos \theta_\mu$ bins) obtained using the same spectral
functions and  vector form factors employed in the calculation of the electron scattering cross section of Fig.~\ref{xsec_ee},
and a dipole parametrization of the axial form factor with $M_A=1.03$~MeV. 

Comparison of Figs. \ref{xsec_ee} and
\ref{dsigma} indicates that the electron and neutrino cross sections corresponding to the same target and 
{\em seemingly} comparable kinematical conditions (the position of the QE peak in Fig.~\ref{xsec_ee} corresponds to
kinetic energy of the scattered electron $\sim~610$~MeV) can not be explained using the same theoretical approach
and the value of the axial mass resulting from deuterium measurements.
In this instance, the paradigm of electron
scattering appears to conspicuously fail.

The above conclusion, while being based on a calculation carried
out within the scheme of Refs.\cite{PRD,NPA}, is largely model independent. Theoretical approaches providing
a quantitative description of the electron-nucleus cross section in the QE channel, are bound to predict CCQE
neutrino-nucleus cross sections significantly below the MiniBooNE data if the value of the axial mass is set to
1.03 GeV. As a matter of fact, within the approach of Refs.\cite{PRD,NPA}, the axial mass yelding
the best $\chi^2$-fit to the flux integrated MiniBooNE $Q^2$-distribution, $M_A=$1.6 GeV, turns out to be even
larger than that reported in Ref.\cite{BooNECCQE}. The authors of Ref.\cite{PRL} have shown that this value
of $M_A$ also explains the muon energy spectrum and angular distribution obtained from integration of the double differential
cross seciton of Ref.\cite{BooNECCQE}.

\begin{figure}[h!]
\includegraphics[width=1.00\columnwidth]{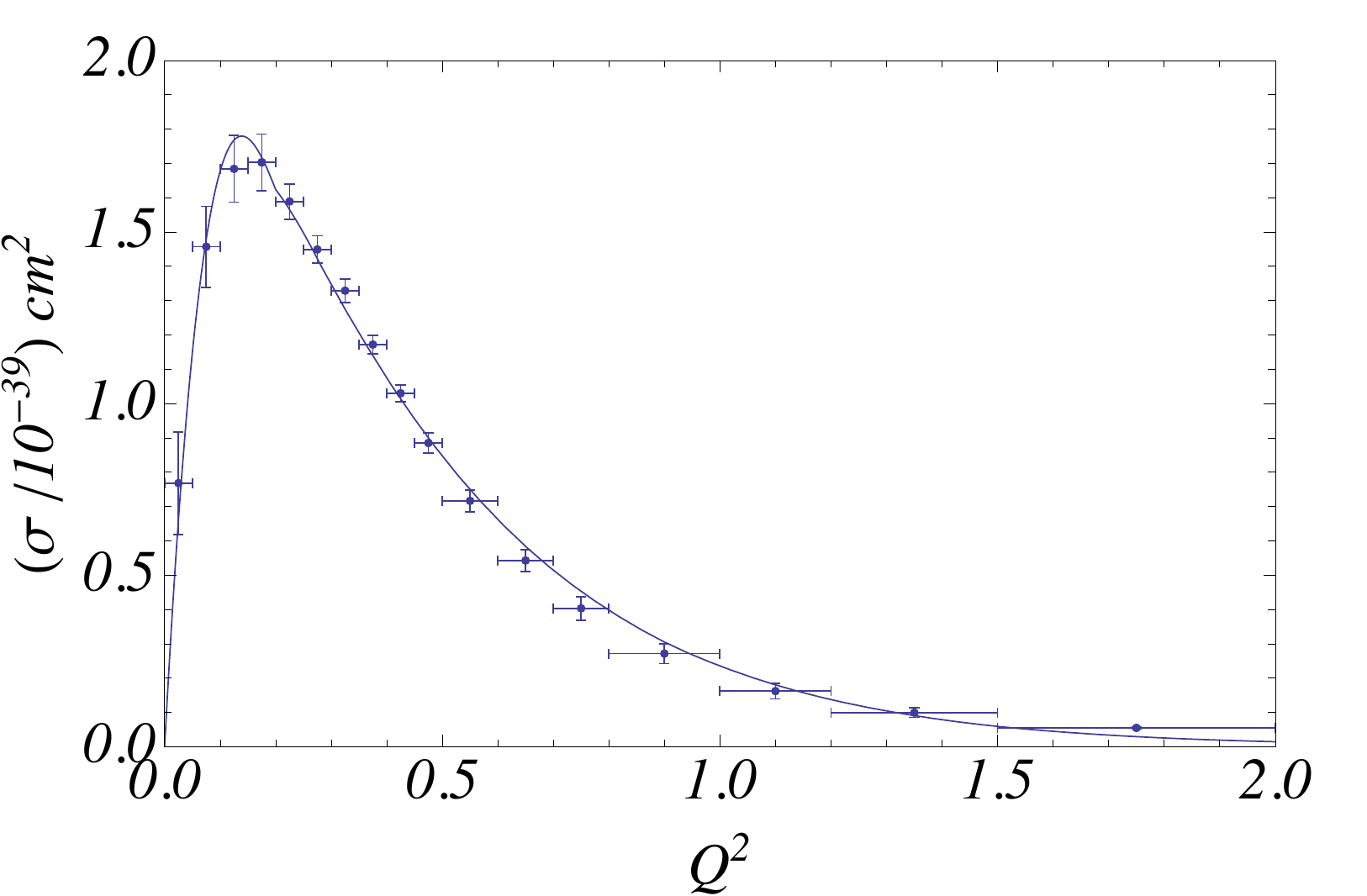}
\caption{Flux averaged $Q^2$-distribution obtained from the approach of Refs.\cite{PRD,NPA} using $M_A=$1.6 GeV, 
compared to the data of Ref.\cite{BooNECCQE}.}
\label{Q2:dist}
\end{figure}

In Fig. \ref{Q2:dist} the theoretical results obtained using $M_A=$1.6 GeV are compared to the distribution of the
 flux-averaged  MiniBooNE events, plotted as a function of the {\em reconstructed} $Q^2$, defined as 
\beq
Q^2 = 2 E_\nu E_\mu \left(1- \frac{p_\mu}{E_\mu} \cos \theta_\mu \right) - m_\mu^2 \  ,
\eeq
where $E_\mu = T_\mu + m_\mu$, $p_\mu = (E_\mu^2 - m_\mu^2)^{1/2}$ and $m_\mu$ is the muon mass.
Note that the definition of $Q^2$, besides the measured kinematical variables $T_\mu$ and $\theta_\mu$, 
involves the incoming neutrino energy $E_\nu$, whose determination necessarily implies some assumptions.

\section{Flux unfolded total cross section}
\label{paradigm}

\begin{figure}[bht]
\includegraphics[width=0.95\columnwidth]{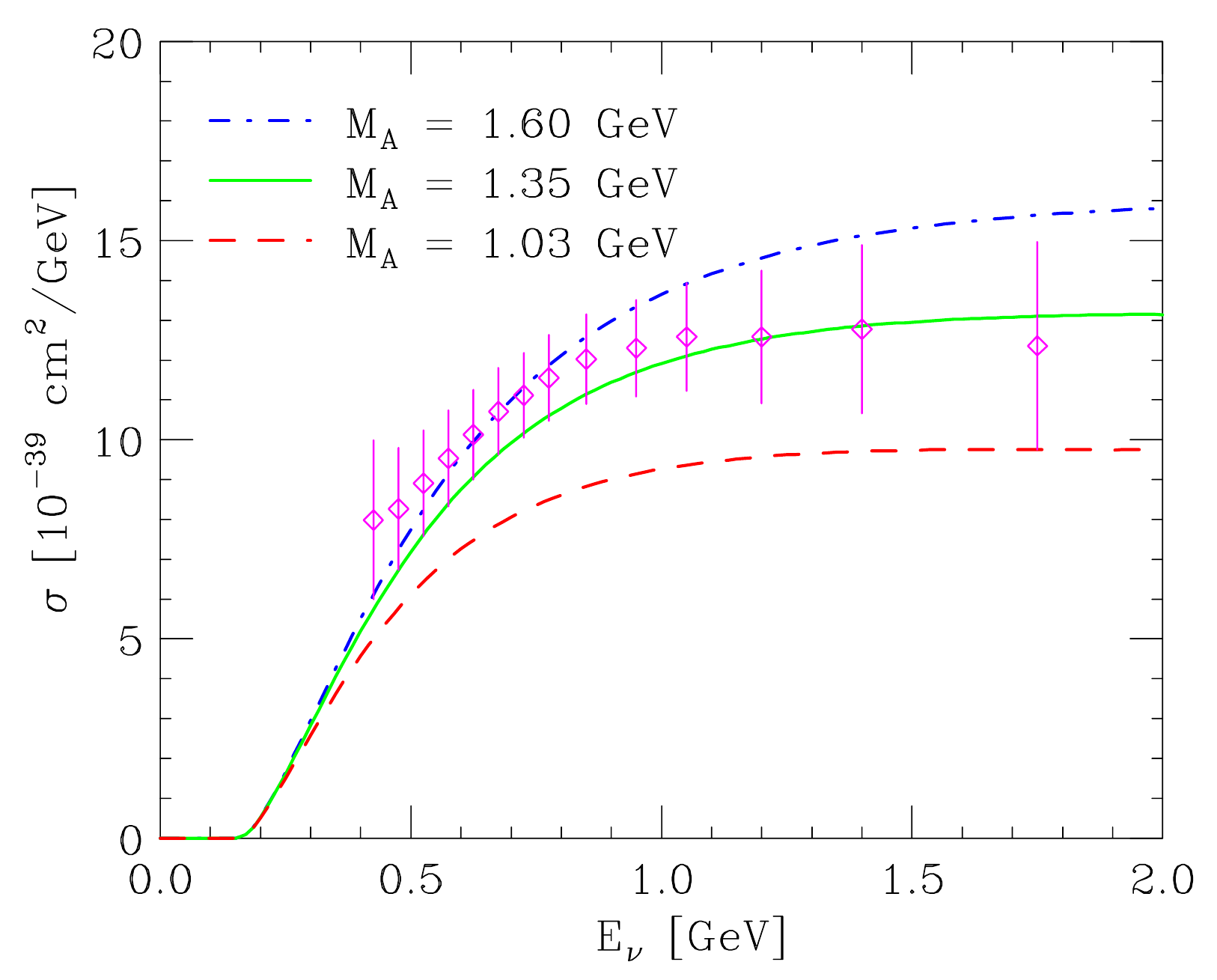}
\caption{Flux unfolded total CCQE cross section, as a function of neutrino energy.
The dot-dash, solid and dashed lines have been obtained setting the value of the axial mass
to $M_A~=~1.03, \ 1.35 \ {\rm and}  \ 1.6$~GeV, respectively. The data are taken from Ref.\cite{BooNECCQE} .}
\label{sigmatot}
\end{figure}

Figure \ref{sigmatot} shows a comparison between the the MiniBooNE {\em flux unfolded} 
total cross section and the results of the calculations of of Ref.\cite{PRL}. It is apparent that in this case 
using $M_A=1.6$ GeV leads to overestimating the data in the region of high energy
($E_\nu > 800$ MeV),
where the choice  $M_A=1.35$ GeV, resulting from he analysis of Ref.\cite{BooNECCQE}, provides a better fit.
The different pattern emerging from Fig. \ref{sigmatot}, compared
to Fig. \ref{Q2:dist}, clearly points to the uncertainty associated with the interpretation of flux averaged and
flux unfolded data.

A different scenario is suggested by the results of Ref.~\cite{martini}, whose authors obtain a
quantitative account of the MiniBooNE flux unfolded total cross section {\em without increasing} $M_A$.
Within the model of Ref.~\cite{martini}, the main mechanism responsible for the enhancement
that brings the theoretical cross section into agreement with the data is multi-nucleon
knock out, leading to $n$ particle-$n$ hole ($n$p-$n$h) nuclear final states ($n=2,3,\ldots$). Within
the approach of Refs.\cite{PRD,NPA}, the occurrence of 2p-2h final
states is described by the continuum part of the spectral function, arising from nucleon-nucleon
correlations \cite{PkE}. It gives rise to the tail extending to large $\omega$, clearly
visible in Fig. \ref{xsec_ee}. However, its contribution turns out to be quite small (less
than 10\% of the integrated spectrum).
The analysis of the momentum distribution sum rule
 indicates that the contributions of $n$p-$n$h final states with $n \geq3$
are negligibly small \cite{BFF}.

According to the philosophy outlined in this paper, in order to firmly establish the role of multi-nucleon
knock out in CCQE neutrino interactions the model of
Ref.~\cite{martini} should be thoroughly tested against electron scattering data.

\section{Summary and outlook}
\label{conclusions}

The theoretical and experimental results discussed in this paper suggest that the main
difference involved in the analysis of neutrino-nucleus scattering, as compared to electron-nucleus scattering, lies in the
flux average.

Unlike the electron cross section shown in Fig. \ref{xsec_ee}, the flux averaged CCQE neutrino cross section
at fixed energy and scattering angle of the outgoing lepton picks up contributions from different kinematical regions, where
different reaction mechanisms dominate. 

\begin{figure}[bht]
\includegraphics[width=0.95\columnwidth]{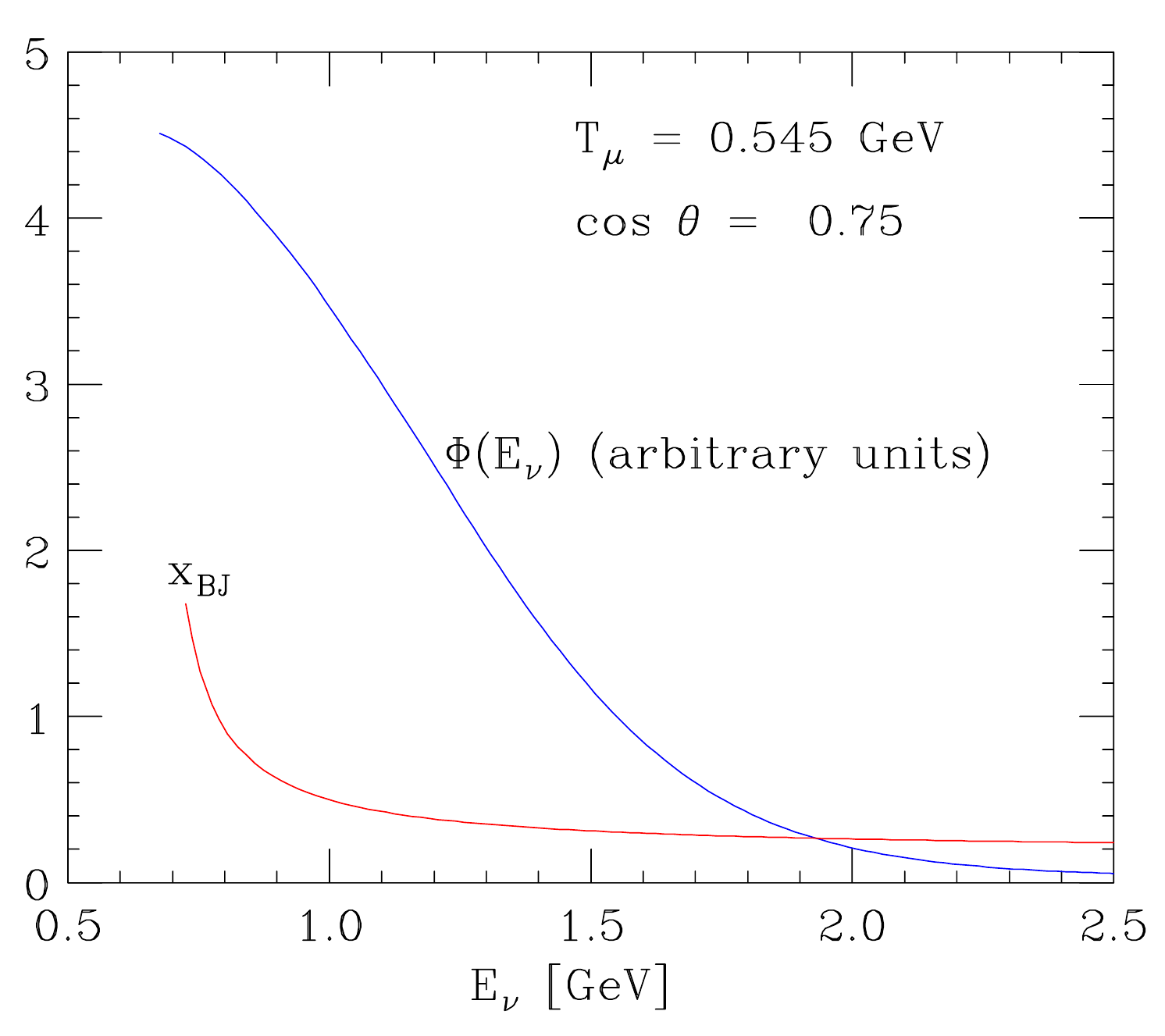}
\caption{Bjorken scaling variable $x$  in the kinematical conditions corresponding to maximum of the cross section 
displayed in the upper panel of Fig. \ref{dsigma}, plotted as a function of the neutrino energy $E_\nu$. For
comparison, the MiniBooNE neutrino flux is also shown.}
\label{columbia}
\end{figure}

Consider, as an example, the cross section at  $\cos\theta_\mu~=~0.75$, the central value of the bin
considered in the upper panel of Fig. \ref{dsigma}, and $T_\mu = 545$ MeV, corresponding to the maximum of
the measured cross section. Figure \ref{columbia} displays the behavior 
of the Bjorken scaling variable $x$ in these kinematical conditions and the MiniBooNe flux as a function of the 
neutrino energy $E_\nu$.

It turns out that $x=1$, corresponding to quasi-elastic kinematics, and $x=0.5$, corresponding to the dip region, 
are associated with neutrino energies $E_\nu~=~788$~MeV (the mean energy of the MiniBooNE flux) and $975$~MeV, respectively, 
 and that 
 \beq
 \frac{  \Phi(E_\nu=975 \ {\rm MeV}) }{ \Phi(E_\nu=788 \ {\rm MeV}) } \approx 0.83 \ 
\eeq
It follows that the flux averaged cross section 
picks up the contributions of quasi-elastic scattering and MEC with about the same probability. Hence, it can not be described according
 to the paradigm successfully applied to electron scattering, based on the tenet that the lepton kinematics is fully determined.

A {\em new paradigm}, suitable for studies of neutrino interactions, should be based on a more flexible model of
nuclear effects, yielding a realistic description of the broad kinematical range associated with the relevant neutrino
energies. 

Nuclear many-body theory provides a consistent framework for the development of such a model. 
Besides single- and multi-nucleon knock out, it should include the contributions of processes involving MEC, 
which are long known to provide a  significant enhancement of the electromagnetic nuclear
response in the transverse channel \cite{euclidean}.  It has to be emphasized that quasielastic scattering and processes involving MEC  
lead to the same final state, and can not be distinguished in the MiniBooNE analysis.
The occurrence of inelastic processes,  leading to excitation of nucleon resonances and pion production should also be taken into account.

As a final remark, it has to be emphasized that a great deal of information could be obtained applying the new paradigm to the analysis 
of {\em inclusive} neutrino-nucleus
cross sections, preferably, although not necessarily, through direct implementation of the resulting nuclear model in the Monte
Carlo simulation codes. This kind of analysis, which has been succesfully carried out for electron-nucleus scattering, may help to 
reconcile the different values of the axial mass obtained by different experiments, and would
be largely unaffected by the problems associated with the possible misidentification of CCQE events, recently
discussed in Ref.~\cite{leitner}.






\section*{Acknowledgements}
This work is based on results obtained in collaboration with P. Coletti and 
D. Meloni. A number of discussions with A.M. Ankowski and C. Mariani on issues related 
to the subject of this paper, are also gratefully acknowledged. 


\end{document}